# ISQ derivation ("derivation in SI units") of a formula for the electrostatic field ionization rate-constant for a hydrogenic atom in its ground electronic state

Richard G. Forbes

Advanced Technology Institute & Department of Electronic Engineering, University of Surrey,

Guildford, Surrey GU2 7XH, UK

Permanent e-mail alias: r.forbes@trinity.cantab.net

As part of their theory, technological applications involving electrostatic field ionization (ESFI), such as gas field ion sources and atom probe tomography, need a formula for the rate-constant $K_e$ for free-space ESFI of a hydrogenic atom in its ground electronic state. This formula needs to explicitly show the dependence on ionization energy $I$ (or, equivalently, the charge-number $Z$). Most existing formulae for hydrogenic atom ESFI were derived in some variant of the atomic units system. However, large numbers of applied scientists and engineers work with ESFI as a process of technological importance, but cannot nowadays be expected to have familiarity with the Gaussian or atomic units equation systems. In the 1970s, what is now called the International System of Quantities (ISQ), which includes the equation system behind SI units, was internationally adopted as the primary system for university teaching and for communication of scientific equations between theoreticians and applied scientists and engineers. 40 years on, derivations of ISQ versions of basic ESFI rate-constant formulae are still not easily found in the literature. Transparent ISQ derivations are now needed. This paper presents a detailed ISQ derivation of a formula for $K_e$, using a method that is modelled closely on the conceptual approach used by Landau and Lifschitz (LL) in their well-known and widely accepted 1958 work (in the atomic units system) on ESFI of the hydrogen atom. This ISQ derivation confirms that, for a hydrogenic atom, the ionization energy appears in the pre-exponential as $I^{5/2}$, and defines a universal "field ionization constant" that also appears. It is also shown how the ISQ formula relates to the Gurney and Condon "attempt frequency" form often used to describe rate-constants for tunnelling processes, and an ISQ expression is given for the motive energy in the related JWKB integral. The derivation involves a motive-energy transformation analogous to a transformation used by LL and in other ESFI papers. The need for such a transformation in ESFI theory raises questions as to the correctness of current theoretical treatments of field electron emission from non-planar emitters, which do not make a transformation of this kind.

# 1. Introduction

## 1.1 Electrostatic and electromagnetic field ionization

Field ionization (FI) is the electric-field-induced ionization of an atom or molecule. In principle, FI can occur either in free space, or at or near the surface of a material, or within a semi-conducting material, or to a molecule within a liquid. This paper is primarily about the theory of *electrostatic field ionization* (ESFI) in free space, but a literature issue requires preliminary comment.

As the name indicates, ESFI results from applying an *electrostatic field*, often the field generated by the charge distributions related to some capacitor-like arrangement. ESFI theory applies even if these charge-distributions are time-varying, as they would be if driven by an alternating voltage or current source, provided that the source-driven oscillations have a time-period very much greater than the time-constant associated with ESFI.

However, logical distinction is needed between ESFI and *electromagnetic FI (EMFI)*, which is field ionization induced by the travelling transverse-electric-field component of an electromagnetic wave. EMFI is particularly relevant when a high-intensity laser beam falls on an atom or molecule. Following early work ([1]; see Ref. [2] for a recent review), conventional EMFI theory is often called *Keldysh theory*.

An established body of thought holds that ESFI is the limiting case of EMFI when the EM radiation frequency becomes very small, and consequently that ESFI theory ought to be the limiting case of EMFI theory. However, this thinking has recently been strongly criticised by Reiss (e.g., [3, 4]), on the grounds (amongst others) that taking ESFI theory as the limiting case of EMFI theory is incompatible with arguments related to Einstein's theory of special relativity.

In the present author's view, the Reiss criticisms are physically credible and need addressing. In the meantime, it seems important to maintain a logical distinction between ESFI theory and EMFI theory, and to accept that ESFI theory can be derived directly from first principles of quantum mechanics. EMFI theory and the relationship between EMFI theory and ESFI theory are outside the scope of this paper.

In reality, some early papers (including some of interest here) that are ostensibly about EMFI assume that, at sufficiently low frequencies, the theory of EMFI is identical with the theory of ESFI, and can be modelled by it. These papers in fact contain theories of ESFI rather than EMFI.

**1.2 Types of electrostatic field ionization**

In the absence of applied fields, an atomic electron is in a state "$\alpha$" of total energy $E_\alpha$. Applying an electrostatic field of magnitude $F$ creates a region, away from the atom, where this electron would have positive kinetic energy. Between this region and the atom's interior there is a potential-energy (PE) barrier, provided the field is not so high that it suppresses the barrier.

Depending on the details, including the values of $E_\alpha$ and $F$, transmission across the barrier could in principle occur either by tunnelling through the barrier or by wave-mechanical "flyover" over it. In practice, either because the field magnitude needs increasing from a low value, or because the atom has to move into a high-field region, ESFI usually occurs by tunnelling before the conditions for flyover are encountered.

A distinction is usefully made between *deep tunnelling* (which occurs well below the barrier peak), and *shallow tunnelling* (which occurs below, but near in energy to, the barrier peak). Because the low-field limit for tunnelling from an atom's ground electronic state is in the deep-tunnelling regime, and also to ensure mathematical validity of quasi-classical arguments used, this paper relates only to the deep tunnelling case.

A further distinction exists between: (a) *near-surface ESFI*, where the PE structure experienced by the removed electron, during removal, is significantly influenced by a nearby material surface; and (b) *free-space ESFI*, where the only significant external influence is the electrostatic field. Obviously, free-space ESFI is the limiting case of near-surface ESFI, as surface-to-atom distance increases.

## 1.3 Technological context

Many technical contexts involve ESFI. (A) Near-surface ESFI generates ions in gas field-ion sources, as used in scanning ion microscopes [5] and potentially other machines, possibly including low-thrust engines for spacecraft. (B) Near-surface ESFI is part of the imaging process in field ion microscopy (FIM) (e.g., [6]). FIM was the first microscopy to "see atoms"; for many years it was an important materials science and surface science technique, but is now mainly an auxiliary technique. (C) In helium FIM, free-space ESFI protects FIM specimens from attack by vacuum-system residual gases, by ionising them in space above the specimen. (D) Free-space ESFI may be involved in electrical breakdown of low-pressure gases. (E) For a positive-polarity pointed needle, in air, free-space ESFI was possibly a partial cause of the "electric wind" phenomenon originally investigated by Priestley in 1766 (see [7]). (F) Free-space ESFI of thermally evaporated neutral atoms may cause unwanted high-energy-deficit ions in liquid metal ion sources [8]. These sources are employed in focused ion beam (FIB) machines, now widely used in many practical and industrial contexts. (G) Near-surface ESFI of field-evaporating metal ions—sometimes called *post-field–ionization (PFI)*—generates higher ion charge states observed in atom-probe tomography (APT), which is a materials science technique of rapidly increasing significance (e.g., [6]).

Processes related to ESFI also occur in dopant-atom ionization in semiconductors, in chemical contexts (where ESFI is called "tunnel ionization"), and in field electron emission from surface states.

The possibility also exists, in atom probe tomography and with liquid metal ion sources, that PFI theory could be used to deduce field values from ion-abundance measurements. No method currently exists for *accurately* measuring fields that vary significantly on a near-atomic scale above a highly-charged surface: a reliable method is urgently needed [9]. PFI theory developed by Kingham [10] in 1982 has been used to estimate fields in this way [11, 12]. Unfortunately, later work [13] found small oversights in Kingham's work, and it is not clear how large any resulting errors are.

The underlying motives for the present work concern this problem of accurate field calibration. Long-term aims are to put PFI theory onto a better and more transparent basis, and assess its accuracy.

However, some preliminary problems need attention.

## 1.4 The need for validation of approximate treatments

Even for the simplest ideal case of near-surface ESFI, which is a hydrogen atom (in its ground electronic state) located close above a simple Sommerfeld model of a flat metal surface, there are no known exact analytical solutions of the Schrödinger equation. Consequently, the main choice is between approximate methods and full numerical solution. Numerical solutions exist for free-space ESFI (see [14]) but not for near-surface-ESFI. Even if practicable, it would probably be laborious, expensive and time-consuming to implement an accurate numerical solution for near-surface ESFI.

A flexible, approximate theoretical approach uses the simple-JWKB (Jeffreys-Wentzel-Kramers-Brillouin) approximation (see below), which derives ultimately from mathematical work by Carlini in 1817 [15] (see [14]). In this approach a one-dimensional *tunnelling probability* $D^{1d}$ is estimated via a JWKB-type integral (see below) taken along an appropriately defined path in space, and the *ESFI rate-constant* $K_e$ is estimated as

$$K_e \approx \nu_e D^{1d} , \tag{1}$$

where $\nu_e$ is the *classical attempt frequency*, identified in simple cases with the classical vibration frequency of an electron in the relevant Bohr-type orbit. This "attempt frequency" formula for tunnelling rate-constants was first discussed by Gurney and Condon in 1928 [17], and is noted by Landau and Lifshitz (LL) in the first English edition of their well-known textbook ([18], eq. (50.10)). Nowadays, the simple-JWKB approximation is also seen as a low-level (but often satisfactory) approximation related to the more general phase-integral method of solving Schrödinger-equation-related and analogous problems [14, 16].

The flexibility of the simple-JWKB method arises because it is straightforward to evaluate the JWKB integral for any well-defined barrier, although numerical integration is usually needed. Following early work by Gomer [19], all treatments of ESFI in FIM/APT theory use this approach (but owe more to earlier field electron emission theory, especially the Burgess et al. 1956 paper [20], than to earlier ESFI theory.)

The issue thus arises of "validating" this simple method against treatments considered more exact, to find the size of the error involved and/or introduce a correction factor. As shown by Haydock and Kingham [21], a plausible approach is to consider ESFI of a hydrogen atom in free space, and compare a derived approximate formula with a more exact treatment, in the low-field limit. In fact, it would be more useful to make the comparison for a hydrogenic atom, of nuclear charge $Ze$, where $e$ is the elementary positive charge.

**1.5 The role of the Landau and Lifshitz rate-constant formula**

The first quantum-mechanical treatment of the free-space ESFI of the hydrogen atom was given by Oppenheimer in 1928 [22], and many later discussions exist. In 1977, Yamabe et al. presented a useful review and new treatments [23], concluding that: "As an exercise in mathematical analysis, the field ionization of the hydrogen atom has exhibited a peculiar perverseness, with unsuspected pitfalls marring some of the earlier calculations. The barrier region dominates practical calculations, which are surprisingly sensitive to the accuracy of the wave-function there". For more recent overviews, see Refs. [14, 24, 25].

Most authorities agree that Landau and Lifshitz (LL) ([18], p. 257) gave the best early treatment. Their result, given in the atomic-units system of measurement and intended to be valid in the low-field limit, was

$$K_\mathrm{e} = (4/F_\mathrm{au})\exp[-2/3F_\mathrm{au}], \tag{2a}$$

where $F_{au}$ is electrostatic field in the atomic units system. This result was originally published in 1958, in the first English edition of their textbook. In the second English edition (1965) ([26], p. 276) LL also gave their result as the Gaussian (unrationalised) system formula:

$$K_e = (4m_e^3 e_s^9 / \hbar^7 F_s) \exp[-2m_e^2 e_s^5 / 3\hbar^4 F_s], \tag{2b}$$

where $m_e$ and $\hbar$ have their usual meanings (see Appendix I). The symbol $e_s$ denotes Gaussian elementary charge, which is related to the ISQ elementary charge $e$ (see below) by $e_s = e/(4\pi\varepsilon_0)^{1/2}$; $F_s$ denotes Gaussian electric field, and is related to ISQ electric field $F$ by $F_s = (4\pi\varepsilon_0)^{1/2}$; $\varepsilon_0$ is the electric constant.

Yamabe et al. [23] reported that a more sophisticated method, using wave-matching techniques, reduced to eq. (2a) in the low-field limit. Thus, the most obvious "first validation method" is to compare simple-JWKB-type formulae with this LL formula (or, rather, its equivalent for a hydrogenic atom). Later, one might wish to make comparisons with higher-order phase-integral methods or other advanced methods.

**1.6 The need for derivations that use the International System of Quantities**

At this point, a serious communication difficulty arises. Almost all fundamental work on free-space ESFI uses atomic units, but many scientists involved in technical applications of ESFI have limited understanding of this system or more generally of (unrationalised) Gaussian equation systems. Thus, many potential users of ESFI theory are not in a good position to follow arguments formulated using these systems.

In the 1970s, the International Standards Organization designated what was then called the "metre-kilogramme-second-ampere system" (or alternatively the "rationalized metre-kilogramme-second system") as the primary equation system for communicating scientific and engineering work.

Since 2009, this system has been formally called the *International System of Quantities (ISQ)* [27, 28]—although, informally, derivations using it are often described as being "in SI units". An important reason for this decision was to facilitate communication between theoreticians and applied scientists and engineers, and the decision had the support of national standards authorities. In many or most places and contexts, the ISQ (which includes the equation system behind SI units) has been the preferred system for undergraduate teaching for the last thirty years or so.

Nevertheless, the theoretical FI community continues to use atomic units. Whilst convenient for certain calculations, this is a serious impediment to mutual understanding between this community and the communities of many hundreds of scientists who use field ionization in working technologies other than those based on laser physics, and the barrier to understanding needs to be diminished.

A particular problem is that it is very difficult to find a derivation of an ISQ equivalent of eq. (2a), either for hydrogen or for a hydrogenic atom (i.e., an atom with one electron but a nuclear charge of $Ze$, where the charge number $Z$ is not necessary integral). The hydrogenic atom, although usually not a realistic model of real atoms and ions, can be a useful basic model for gaining understanding about the ESFI of real atoms and ions. The present paper aims to fill this gap, and also to indicate the relationship between ISQ equivalents of eq. (2a) and related "attempt-frequency-type" formulae.

The primary aim here is provide an ISQ derivation of a suitable basic formula. The original LL treatment puts the ionization energy of the hydrogen ground state equal to ½. In free space this may be convenient. However, in near-surface ESFI there can be significant shifts in the effective ionization energy, due to image forces, and what one wishes to know (as a starting point) is the power to which the ionization energy $I$ for a hydrogenic atom is raised in the pre-exponential of the rate-constant formula for *free-space* ESFI from the ground electronic state. Several sources, e.g. Refs [29-31], imply that the power is $I^{5/2}$. A secondary aim here is to confirm that that an ISQ treatment modelled closely on the LL treatment of hydrogen also reaches this result.

For completeness, one should add that treatments (e.g., Refs [32-34]) of more realistic atomic models appear to yield results different from $I^{5/2}$, but what one needs to know for an initial re-examination of current models of post-field-ionization in atom probe tomography is the behaviour of the hydrogenic atom.

A derivation of ESFI rate-constant could be provided here for excited electron states, but reality is that many technical applications involve ESFI from the ground state. For simplicity, the treatment here relates to the ground state only. The method used is modelled closely on the conceptual approach used by Landau and Lifschitz. Relativistic effects and effects related to hyperfine splitting are disregarded, as in the LL treatment.

The structure of this tutorial paper is as follows. Section 2 sets out some theoretical preliminaries; Section 3 presents the ISQ derivation of an LL-type rate-constant formula for a hydrogenic atom; Section 4 puts this formula into alternative forms; and Section 5 presents a summary and discussion. Appendix I deals with the definitions and values of constants, and Appendix II relates formula (2b) to results derived below.

## 2. Some detailed theoretical background

Formulae are given in ISQ form unless otherwise indicated. To reduce algebraic complexity, defined constants are used to represent combinations of fundamental constants and (in some cases) problem-specific constants. Values are given both in SI units and in the ISQ-compatible units (based on the eV and the V/nm) often used in field electron emission and in practical applications of field ion emission (see [35]). ISQ formulae can, as usual, be converted to atomic units form by allocating appropriate numerical values to the constants. Appendix I gives details.

### 2.1 Schrödinger-equation issues

In one-electron theory, the three-dimensional Schrödinger equation can be written

$$(\hbar^2/2m^*)\Delta\Psi + (E-U)\Psi = 0, \tag{3}$$

where $\hbar$ is Planck's constant divided by $2\pi$, $m^*$ the relevant electron mass, $U$ the electron PE, $E$ the total electron energy (relative to the same reference zero as $U$), and $\Delta$ the Laplacian operator. Strictly, for a hydrogenic atom, $m^*$ should be taken as the reduced mass $\mu_e$ of an electron moving around the nucleus (e.g., [36]). However, the difference between $\mu_e$ and the free-space electron mass $m_e$ is very small, and the present treatment can neglect it, by replacing $m^*$ by $m_e$.

On introducing the parameter $\sigma$ (which is Fowler & Nordheim's parameter $\kappa$ [37]), defined by

$$\sigma \equiv (2m_e)^{1/2}/\hbar \tag{4}$$

(see Appendix I), the one-dimensional Schrödinger equation for motion along some Cartesian coordinate $\ell$ can be written

$$\partial^2 \psi_\ell / \partial \ell^2 \;=\; \sigma^2 (U_\ell - E_\ell) \psi_\ell(\ell) \;\equiv\; \sigma^2 M(\ell) \psi_\ell(\ell), \tag{5}$$

where $\psi_\ell(\ell)$ is the related wave-function (or wave-function component), $U_\ell$ the effective potential energy for motion along the coordinate, and $E_\ell$ the effective total energy (or total-energy component). The quantity $M(\ell)$ [$\equiv U_\ell - E_\ell$], sometimes called the *motive energy*, is defined by eq. (5). A finite range of $\ell$ where $M(\ell) \geq 0$ constitutes a tunnelling barrier for an electron of energy $E_\ell$; $M(\ell)$ describes the mathematical form of the barrier.

If $\ell_{in}$ and $\ell_{out}$ are relevant zeros of $M(\ell)$, at either end of the barrier, then a parameter $G$ called here the *barrier strength* (also known as the "Gamow exponent" and the "JWKB exponent") is defined by

$$G = 2\sigma \int_{\ell_{in}}^{\ell_{out}} M^{1/2}(\ell) \mathrm{d}\ell . \tag{6}$$

This parameter $G$ characterises the barrier. The *Landau & Lifshitz tunnelling probability formula*,

which applies to a one-dimensional treatment of deep tunnelling (see [38, 39] and eq. (50.12) in [18]) takes the tunnelling probability $D$ for this one-dimensional barrier as given by:

$$D = P_t \exp[-G], \tag{7}$$

where $P_t$ is a *tunnelling pre-factor*. This factor $P_t$ is usually of order unity but is difficult to calculate precisely. The *simple-JWKB approximation* sets $P_t=1$, and takes $D \approx \exp[-G]$. The above represents the usual treatment of deep tunnelling found in literature other than FI literature.

### 2.2 Coordinate-system issues

For the free-space ESFI of a hydrogenic atom, the Schrödinger equation separates in parabolic coordinates; hence these are used in exact treatments. Several types exist. That used by LL (see §37 of [18] or [26]) relates Cartesian coordinates $\{x,y,z\}$ to parabolic coordinates $\{\xi,\eta,\phi\}$ by

$$x = \xi^{1/2}\eta^{1/2}\cos\phi; \quad y = \xi^{1/2}\eta^{1/2}\sin\phi; \quad z = (\xi - \eta)/2, \tag{8}$$

$$r = (x^2 + y^2 + z^2)^{1/2} = (\xi + \eta)/2, \tag{9}$$

with the Laplacian operator $\Delta$ given by

$$\Delta = \frac{4}{\xi+\eta}\left[\frac{\partial}{\partial\xi}\left(\xi\frac{\partial}{\partial\xi}\right) + \frac{\partial}{\partial\eta}\left(\eta\frac{\partial}{\partial\eta}\right)\right] + \frac{1}{\xi\eta}\frac{\partial^2}{\partial\phi^2} \ . \tag{10}$$

For the present author's longer-term objectives, there is a problem with this system. If one applies to a hydrogenic atom an electrostatic field that (in Cartesian coordinates) is positive in the positive *z*- direction, then the electron leaves the atom in the negative *z*-direction in Cartesian coordinates, but in the positive $\eta$-direction in parabolic coordinates. This minus sign is potentially confusing when comparing treatments in parabolic and Cartesian coordinates, and is a serious inconvenience when discussing near-surface ESFI, where it is much more transparent (when using Cartesian coordinates) to take the electrostatic field as negative, and then evaluate integrals in the positive *z*-direction.

In parabolic coordinates as defined above, if one takes the field as negative in the Cartesian positive *z*-direction, then in parabolic coordinates the electron leaves the atom in the positive $\xi$-direction. In this case, the formulae derived in an LL-type treatment end up expressed in terms of $\xi$ rather than $\eta$, and this also is potentially confusing.

A solution is to exchange the roles of $\eta$ and $\xi$, so that definitions (8) become

$$x = \eta^{1/2}\xi^{1/2}\cos\phi; \quad y = \eta^{1/2}\xi^{1/2}\sin\phi; \quad z = (\eta - \xi)/2 , \tag{11}$$

but eqns (9) and (10) remain unchanged. The electron then leaves the atom in the positive *z*-direction in Cartesian coordinates and the positive $\eta$-direction in parabolic coordinates. These modified coordinates are used below.

### 2.3 Basic ISQ results for a hydrogenic atom, in the zero-field situation

For an electron in a hydrogenic atom (HA), the potential energy *U* (relative to infinity), at distance *r* from the nucleus, can be written

$$U = -Ze^2/4\pi\varepsilon_0 r \equiv -B/r, \tag{12}$$

where $\varepsilon_0$ is the electric constant, and *B* [$\equiv Ze^2/4\pi\varepsilon_0$] is defined by eq. (12).

When the electron's reduced mass is set equal to its free-space mass, the wave-function $\Psi$ of the

HA ground electronic state is (e.g., [36])

$$\Psi = \frac{Z^{3/2}}{\pi^{1/2} a_0^{3/2}} \exp[-Zr/a_0] = \frac{1}{\pi^{1/2} a_Z^{3/2}} \exp[-r/a_Z], \qquad (13)$$

where (in the ISQ) the *Bohr radius* $a_0$ is given by (e.g., [36])

$$a_0 = 4\pi\varepsilon_0 \hbar^2 / e^2 m_e = (2/\sigma^2)(4\pi\varepsilon_0 / e^2), \qquad (14)$$

and the classical radius $a_Z$ of the ground-state Bohr-type orbit of an HA is

$$a_Z = a_0/Z = 2/\sigma^2 B. \qquad (15)$$

The HA ionization energy $I$ can be written in any of the equivalent forms (see [36, 40])

$$I = Z^2 e^4 m_e / 32\pi^2 \varepsilon_0^2 \hbar^2 = Z^2 I_H, \qquad (16a)$$

$$I = B/2a_Z = B^2 \sigma^2 / 4 = \tfrac{1}{2} m_e c_Z^2, \qquad (16b)$$

where $I_H$ is the ionization energy of a hydrogen atom, and $c_Z$ is the classical electron velocity in the lowest HA Bohr-type orbit. This velocity is determined by the Bohr quantisation condition $m_e c_Z a_Z = \hbar$, (e.g., [40]). Hence the *classical orbital vibration frequency* $\nu_Z$ is given by

$$\nu_Z = c_Z / 2\pi a_Z = \frac{m_e c_Z^2}{2\pi\hbar} = \frac{I}{\pi\hbar} = \frac{Z^2 I_H}{\pi\hbar}. \qquad (17)$$

From eqns (13) and (16b), $1/a_Z = 2I/B$, and the HA ground-state wave-function (in zero field) can

be written

$$\Psi = \pi^{-1/2}(2I/B)^{3/2}\exp[-(2I/B)\cdot r],\qquad(18)$$

In parabolic coordinates, of either type discussed above, eq. (12) and the related Schrödinger equation become

$$U = -2B/(\eta+\xi),\qquad(19)$$

$$\frac{4}{\eta+\xi}\left[\frac{\partial}{\partial\eta}\left(\eta\frac{\partial\Psi}{\partial\eta}\right)+\frac{\partial}{\partial\xi}\left(\xi\frac{\partial\Psi}{\partial\xi}\right)\right]+\frac{1}{\eta\xi}\frac{\partial^2\Psi}{\partial\phi^2}+\sigma^2\left(E+\frac{2B}{\eta+\xi}\right)\Psi = 0.\qquad(20)$$

As usual when separating variables in cylindrically symmetric situations, one looks for solutions of the form

$$\Psi(\eta,\xi,\phi) = \psi_\eta\psi_\xi\psi_\phi = \psi_\eta\psi_\xi\cdot\lambda_\phi e^{im\phi} = \psi_\eta\psi_\xi\cdot(2\pi)^{-1/2}e^{im\phi},\qquad(21)$$

where $\{\psi_\eta,\psi_\xi,\psi_\phi\}$ are the separated wave-function components, considered as separately normalised, and $m$ is the magnetic quantum number. The parameter $\lambda_\phi$ is the normalisation constant for the $\phi$-coordinate, and has the value $(2\pi)^{-1/2}$.

Substituting eq. (21) into eq. (20), and then evaluating $\partial^2\psi_\phi/\partial\phi^2$, dividing by $\Psi$, and multiplying by $(\eta+\xi)/4$, yields

$$\frac{1}{\psi_\eta}\frac{\partial}{\partial\eta}\left(\eta\frac{\partial\psi_\eta}{\partial\eta}\right)+\frac{1}{\psi_\xi}\frac{\partial}{\partial\xi}\left(\xi\frac{\partial\psi_\xi}{\partial\xi}\right)+-m^2\left(\frac{1}{4\eta}+\frac{1}{4\xi}\right)+\sigma^2\left(\frac{E\eta}{4}+\frac{E\xi}{4}+\frac{B}{2}\right)=0\qquad(22)$$

Separating variables gives (where $\beta_\eta$ and $\beta_\xi$ are separation constants):

$$\frac{1}{\psi_\eta}\frac{\partial}{\partial \eta}\left(\eta\frac{\partial \psi_\eta}{\partial \eta}\right) - \frac{m^2}{4\eta} + \frac{\sigma^2 E\eta}{4} + \beta_\eta = 0, \tag{23}$$

$$\frac{1}{\psi_\xi}\frac{\partial}{\partial \xi}\left(\xi\frac{\partial \psi_\xi}{\partial \xi}\right) - \frac{m^2}{4\xi} + \frac{\sigma^2 E\xi}{4} + \beta_\xi = 0. \tag{24}$$

$$\beta_\xi + \beta_\eta = \sigma^2 B/2. \tag{25}$$

In the ground electronic state, $m=0$, the total energy $E$ is $-I$, and the wave-function is symmetric. Hence $\beta_\eta = \beta_\xi = \sigma^2 B/4$, and the separated equations for $\eta$ and $\xi$ reduce to

$$\frac{1}{\psi_\eta}\frac{\partial}{\partial \eta}\left(\eta\frac{\partial \psi_\eta}{\partial \eta}\right) + \sigma^2\left(\frac{B}{4} - \frac{I\eta}{4}\right) = 0, \tag{26}$$

$$\frac{1}{\psi_\xi}\frac{\partial}{\partial \xi}\left(\xi\frac{\partial \psi_\xi}{\partial \xi}\right) + \sigma^2\left(\frac{B}{4} - \frac{I\xi}{4}\right) = 0. \tag{27}$$

In parabolic coordinates, eq. (18) becomes

$$\Psi = \pi^{-1/2}(2I/B)^{3/2}\exp[-(I/B)\cdot(\eta+\xi)], \tag{28}$$

and clearly separates into the components of the form

$$\psi_\eta = \lambda_\eta \exp[-(I/B)\eta]; \quad \psi_\xi = \lambda_\xi \exp[-(I/B)\xi]; \quad \psi_\phi = (2\pi)^{-1/2}. \tag{29}$$

Here, $\lambda_\eta$ and $\lambda_\xi$ are normalisation constants for the $\eta$ and $\xi$ coordinates. Individual values for $\lambda_\eta$ and are not needed in LL's derivation. It is readily confirmed, using eq. (16), that $\psi_\eta$ and $\psi_\xi$, as given by eq. (29), are solutions of eqns (26) and (27) respectively.

**2.4 Quasi-classical wave-functions**

The use of quasi-classical wave-functions is a standard technique described in many quantum mechanics textbooks, including LL (see §50 of [18] or [26]). LL write relevant formulae basically as follows. For electron motion along a Cartesian coordinate, here denoted by $\ell$, where a barrier exists, let the outer classical turning point for the barrier be denoted by $\ell_1$, and let the function $p(\ell)$ be

$$p(\ell) = +\sqrt{2m_e \{E_\ell - U_\ell(\ell)\}} \ . \tag{30}$$

The required quasi-classical wave-functions are

$$\text{for } \ell < \ell_1: \quad \psi_\ell = -\frac{iC_n}{\sqrt{|p|}} \exp\left[\hbar^{-1} | \int_{\ell_1}^{\ell} p\, d\ell \,|\right] , \tag{31}$$

$$\text{for } \ell > \ell_1: \quad \psi_\ell = \frac{C_n}{\sqrt{p}} \exp\left[\{i\hbar^{-1} \int_{\ell_1}^{\ell} p\, d\ell\} - i\pi/4\right]. \tag{32}$$

where $C_n$ is a normalisation constant. Apart from small notation differences, these equations correspond to LL, eqns. (50.4) of [18] or [26].

**3. Derivation of a formula for the ESFI rate-constant for a hydrogenic atom**

## 3.1 Including the electrostatic field in the Schrödinger equation

The Schrödinger equation must next be modified to include an applied electrostatic field. As discussed above, the approach here differs slightly from that of LL, but yields the same mathematical equation in terms of $\eta$.

In Cartesian coordinates, if an electrostatic field $\mathcal{E}$ is applied in the positive $z$-direction, the PE formula (12) for a HA is replaced by

$$U = -B/r + e\mathcal{E}z \ . \tag{33}$$

Choose $\mathcal{E}$ to be negative, and let $F$ denote its magnitude. Thus, $\mathcal{E} = -F$, and eq. (33) becomes

$$U = -B/r - eFz \ . \tag{34}$$

In the modified parabolic coordinates defined above, this becomes

$$U = -2B/(\eta+\xi) - eF(\eta-\xi)/2 \ . \tag{35}$$

Hence, on applying the same procedures as before, eq. (16) becomes replaced by

$$\left[\frac{1}{\psi_\eta}\frac{\partial}{\partial\eta}\left(\eta\frac{\partial\psi}{\partial\eta}\right) + \frac{1}{\psi_\xi}\frac{\partial}{\partial\xi}\left(\xi\frac{\partial\psi_\xi}{\partial\xi}\right)\right] - \frac{m^2}{4\eta} - \frac{m^2}{4\xi} + \sigma^2\left[\frac{-I(\eta+\xi)}{4} + \frac{eF(\eta^2-\xi^2)}{8} + \frac{B}{2}\right] = 0 \ , \tag{36}$$

where $I$ here represents the ionization energy of the polarised HA, which is greater than that of the unpolarised HA by the Stark shift.

For all field values of interest, the Stark shift is small compared with the ionization energy of the unpolarised HA. Further, since LL needed a formula valid in the low-field limit, they did not consider details of the shift. This approach, effectively of neglecting the Stark shift, is also adopted here.

As before, eq. (36) separates, giving, for the ground-state (for which $m=0$):

$$\frac{1}{\psi_\eta}\frac{\partial}{\partial\eta}\left(\eta\frac{\partial\psi_\eta}{\partial\eta}\right)+\sigma^2\left[\frac{-I\eta}{4}+\frac{eF\eta^2}{8}\right]+\beta_\eta=0 \qquad (37)$$

$$\frac{1}{\psi_\xi}\frac{\partial}{\partial\xi}\left(\xi\frac{\partial\psi_\xi}{\partial\xi}\right)+\sigma^2\left[\frac{-I\xi}{4}-\frac{eF\xi^2}{8}\right]+\beta_\xi=0 , \qquad (38)$$

$$\beta_\eta+\beta_\xi=\sigma^2 B/2 . \qquad (39)$$

With a field present, $\beta_\eta$ and $\beta_\xi$ are no longer equal, but may be written ([18] §73 or [26] §77)

$$\beta_\eta=(\sigma^2 B/4)+\delta\beta_\eta, \quad \beta_\xi=(\sigma^2 B/4)+\delta\beta_\xi, \qquad (40)$$

where $\delta\beta_\eta$ and $\delta\beta_\xi$ can be written as functions of $F$.

## 3.2 Derivation of an equation for $|\chi_\eta|^2$

LL applied quasi-classical arguments to the coordinate $\eta$. To get eq. (37) into a suitable form, analogous to the Cartesian one-dimensional Schrödinger equation, they substitute

$$\psi_\eta=\eta^{-1/2}\chi_\eta, \qquad (41)$$

which yields

$$\frac{\eta}{\chi_\eta}\frac{\partial^2 \chi_\eta}{\partial \eta^2}+\frac{1}{4\eta}+(\sigma^2/2)\left[-\frac{I\eta}{2}+\frac{B}{2}+\frac{eF\eta^2}{4}\right]+\delta\beta_\eta = 0 \ . \tag{42}$$

When $F$ becomes small, $\delta\beta_\eta$ also becomes small; hence LL disregard this term, and eq. (42) yields

$$\frac{\partial^2 \chi_\eta}{\partial \eta^2}+(\sigma^2/2)\left[-\frac{I}{2}+\frac{B}{2\eta}+\frac{1}{2\sigma^2\eta^2}+\frac{eF\eta}{4}\right]\chi_\eta = 0 \ . \tag{43}$$

To apply this equation to the hydrogen atom, in the low-field limit, using atomic units, put $e=\rightarrow 1$, $\sigma^2 \rightarrow 2$, $B \rightarrow 1$, $I \rightarrow \frac{1}{2}$, yielding

$$\frac{\partial^2 \chi_\eta}{\partial \eta^2}+\left[-\frac{1}{4}+\frac{1}{2\eta}+\frac{1}{4\eta^2}+\frac{F\eta}{4}\right]\chi_\eta = 0 \ . \tag{44}$$

Effectively, this is eq. (2) on p. 257 of [18] (p. 275 of [26]), with $F$ taking the place of $\mathcal{E}$.

Equation (43) can also be written as the linked equations

$$\frac{\partial^2 \chi_\eta}{\partial \eta^2}=\sigma^2 \mathfrak{M}(\eta) \ , \tag{45}$$

$$\mathfrak{M}(\eta) = \frac{I}{4}-\frac{eF\eta}{8}-\frac{B}{4\eta}-\frac{1}{4\sigma^2\eta^2} \ , \tag{46}$$

where, in respect of the transformed Schrödinger equation (43), $\mathfrak{M}(\eta)$ plays the mathematical role of the function $M(\ell)$ defined earlier. By analogy with eq. (30), a function $p(\eta)$ is related to $\mathfrak{M}(\eta)$ as follows. Inside the barrier, $p(\eta)$ is purely imaginary, with $p/i$ positive and

$$|p(\eta)| = +\sqrt{2m_e \mathcal{M}(\eta)} \ . \tag{47}$$

Outside the barrier, $p(\eta)$ is real and positive with

$$p(\eta) = +\sqrt{-2m_e \mathcal{M}(\eta)} \ . \tag{48}$$

In the context of eq. (43), this parameter $p$ plays the mathematical role of the parameter $p$ in eqns (30) to (32). To avoid confusion, a different typeface is used here because, in the context of eq. (43), $p$ is not a Cartesian quantum-mechanical momentum.

### 3.3 Derivation of a quasi-classical expression for $|\Psi|^2$

The physics is now treated in a manner equivalent to LL's treatment. Both inside and outside the barrier (i.e., for large $\eta$), the full three-dimensional wave-function can be written

$$X(\eta,\xi,\phi) = \chi_\eta \psi_\xi \psi_\phi = \eta^{1/2} \psi_\eta \psi_\xi \psi_\phi = \eta^{1/2} \Psi(\eta,\xi,\phi) \ . \tag{49}$$

Because the Schrödinger equation has been separated, $\psi_\phi$ and $\psi_\xi$ are the same inside and outside the barrier, and quasi-classical arguments can be applied to relate the values of $\chi_\eta$ (and hence $X$ and $\Psi$) inside and outside the barrier.

LL define $\eta_0$ to be some value of $\eta$ "within the barrier", requiring that (in atomic units) $1 \ll (\eta_0)_{au} \ll 1/F_{au}$. In the ISQ, for an HA, the equivalent condition appears to be $a_Z \ll \eta_0 \ll (2I/eF)$, but precise numerical limits are not important. What is required physically is that $\eta_0$ be "firmly within the barrier, but as close to the inner edge of the barrier as the validity of quasi-classical arguments permits". LL also define $\eta_1$ to be the barrier's outer classical turning point.

Quasi-classical arguments are in principle mathematical arguments about the properties of a particular class of second-order ordinary differential equations. They may be applied to eq. (43) for $\chi_\eta(\eta)$, because this equation has the mathematical form of the Cartesian one-dimensional Schrödinger equation. Thus, the quasi-classical mathematical treatment applied by LL earlier in their book can be applied here. The procedure uses eq. (31) (the quasi-classical expression for the wave-function within the barrier) to find $C_n$, by comparing eq. (31) with the assumed exact (or "nearly exact") wave-function at $\eta_0$. The wave-function outside the barrier is then given by eq. (32).

LL assume that, at position $(\eta_0,\xi,\phi)$, relatively close to the nucleus, the wave-function can be approximated as that of the unpolarised atom. Consequently, from eqns (28) and (49), the value of $X(\eta_0,\xi,\phi)$ is:

$$X(\eta_0,\xi,\phi) = \pi^{-1/2}(2I/B)^{3/2}\eta_0^{1/2}\exp[-(I/B)\cdot(\eta_0+\xi)]. \tag{50}$$

From eq. (31), with $\ell$ replaced by $\eta$, and $p$ by $p$, the quasi-classical expression for $X(\eta_0,\xi,\phi)$ is

$$X(\eta_0,\xi,\phi) = -\frac{iC_n(\xi,\phi)}{\sqrt{|p_0|}}\exp\{\hbar^{-1}|\int_{\eta_1}^{\eta_0} p\,d\eta|\} = +\frac{iC_n(\xi,\phi)}{\sqrt{|p_0|}}\exp\{\hbar^{-1}\int_{\eta_0}^{\eta_1}|p|\,d\eta\}, \tag{51}$$

where (inside the barrier) $|p(\eta)| = +\sqrt{2m_e\mathfrak{M}(\eta)}$, as above, and $p_0 = p(\eta_0)$. In the second expression, the order of integration has been reversed. Also, because, inside the barrier, $p$ is purely imaginary and $p/i$ positive, the modulus signs around the integral have been re-positioned. Equating the two expressions for $X(\eta_0,\xi,\phi)$, and also using eq. (47) and definition (4), yields

$$C_n(\xi,\phi) = \frac{\sqrt{|p_0|}\cdot\{\pi^{-1/2}(2I/B)^{3/2}\eta_0^{1/2}\}\cdot\exp[-(I/B)\cdot(\eta_0+\xi)]}{i\exp[\sigma\int_{\eta_0}^{\eta_1}\mathfrak{M}^{1/2}(\eta)d\eta]}. \tag{52}$$

Inserting this value into eq. (32) (with $\ell$ replaced by $\eta$, and $p$ by $\wp$), yields, for the wave-function at position $(\eta,\xi,\phi)$ somewhat outside the barrier:

$$X(\eta,\xi,\phi) = \pi^{-1/2}(2I/B)^{3/2}\} \cdot \eta_0^{1/2} \exp[-(I/B)(\eta_0+\xi)] \cdot \frac{\sqrt{|\wp_0|}}{\sqrt{\wp}} \frac{\exp\left[i\{\hbar^{-1}|\int_{\eta_1}^{\eta}\wp d\eta|\} - 3i\pi/4\right]}{\exp\left[\sigma \int_{\eta_0}^{\eta_1} \mathfrak{M}^{1/2}(\eta)d\eta\right]}, \quad (53)$$

where the "1/i" term in eq. (52) has been converted to "exp(–i$\pi$/2)" and has contributed to the term "exp(–i3$\pi$/4)" in eq. (53).

LL now introduce relatively crude approximations for $|\wp_0|$ and $\wp$. That for $|\wp_0|$ (inside the barrier) is derived by selecting only the first term in eq. (46); that for $\wp$ (outside the barrier) is derived by selecting only the first two terms in eq. (46). This yields, using eqns (47) and (48):

$$\frac{|\wp_0|}{\wp} \approx \frac{\sqrt{(2m_e)(I/4)}}{\sqrt{(2m_e)(eF\eta/8 - I/4)}} = \frac{1}{\sqrt{eF\eta/2I - 1}}. \quad (54)$$

Hence, at coordinate-value $\eta$ outside the barrier, the squared modulus of the wave-function can be written as the set of linked equations

$$|X(\eta,\xi,\phi)|^2 = \gamma(\eta,\xi,\phi) \cdot T, \quad (55a)$$

$$\gamma(\eta,\xi,\phi) \equiv \pi^{-1} \cdot (2I/B)^2 \cdot (eF\eta/2I - 1)^{-1/2} \cdot \exp[-(2I/B)\xi], \quad (55b)$$

$$T \equiv (2I/B)\eta_0 \exp[-(2I/B)\eta_0] \cdot \exp[-g], \quad (55c)$$

$$g \equiv 2\sigma \int_{\eta_0}^{\eta_1} \mathfrak{M}^{1/2}(\eta)d\eta. \quad (55d)$$

Here, $\mathfrak{M}(\eta)$ is as given by eq. (46), $T$ is a "barrier term" that is the same for all values of $\phi$ and $\xi$ ($\xi$>0), and $\gamma(\eta,\xi,\phi)$ is a "geometrical term" that has to be integrated over all relevant values of $\phi$ and $\xi$. The pre-exponential term $(2I/B)$ [=$1/a_Z$] is included in eq. (55c) to make $T$ dimensionless, and a compensating term has been removed when writing eq. (55b).

Using eq. (46) in eq. (55d) yields

$$g = \int_{\eta_0}^{\eta_1} \sigma \left( I - \frac{eF\eta}{2} - \frac{B}{\eta} - \frac{1}{\sigma^2 \eta^2} \right)^{1/2} d\eta. \tag{56}$$

LL now use an approximation that takes the first three terms in the bracket in eq. (56) and (neglecting the fourth term) binomially expands them in the form

$$\sigma I^{1/2} \left[ 1 - \frac{eF\eta}{2I} - \frac{B}{\eta I} \right]^{1/2} \approx \sigma I^{1/2} \left[ \left\{ 1 - \frac{eF\eta}{2I} \right\}^{1/2} - \left( \frac{1}{2} \frac{B}{\eta I} \right) \left\{ 1 - \frac{eF\eta}{2I} \right\}^{-1/2} \right]. \tag{57}$$

The first term in the expanded expression integrates straightforwardly to give

$$g_1 = \left( \frac{4\sigma}{3e} \right) \frac{I^{3/2}}{F} \left[ -\left\{ 1 - \frac{eF\eta_1}{2I} \right\}^{3/2} + \left\{ 1 - \frac{eF\eta_0}{2I} \right\}^{3/2} \right]. \tag{58}$$

This is simplified as follows. By selecting (and equating to zero) the first two terms in expression (46), one can approximate $\eta_1 \approx 2I/eF$, which reduces the first term in eq. (58) to approximately zero. For the second term, take ($eF\eta_0/2I$) as <<1, and expand binomially to yield

$$g_1 \approx \left[ \left( \frac{4\sigma}{3e} \right) \frac{I^{3/2}}{F} \right] \cdot \left\{ 1 - \frac{3eF\eta_0}{4I} \right\} = [bI^{3/2}/F] - [\sigma I^{1/2} \eta_0], \tag{59}$$

where $b$ [$\equiv 4\sigma/3e$] is an universal constant well known in field electron emission and sometimes called the *second Fowler-Nordheim constant* (see Appendix I).

Integration of the final term in eq. (57), by substituting $t=(1-eF\eta/2I)^{1/2}$, and then using partial fractions, yields a contribution

$$g_2 = (\sigma B/2I^{1/2}) \cdot \left[ \ln\left\{\frac{1+(1-eF\eta_1/2I)^{1/2}}{1-(1-eF\eta_1/2I)^{1/2}}\right\} - \ln\left\{\frac{1+(1-eF\eta_0/2I)^{1/2}}{1-(1-eF\eta_0/2I)^{1/2}}\right\} \right]. \tag{60}$$

This expression is simplified as follows. Equation (16b) shows that $\sigma B/2I^{1/2} = 1$. Then, in the first logarithm, $(1-eF\eta_1/2I)$ is treated as small, so the outer bracket reduces to approximately unity and the logarithm to approximately zero. In the second logarithm, $(eF\eta_0/2I)$ is treated as small, so binomial expansions yield

$$g_2 \approx -\ln\{\eta_0^{-1}(8I/eF)\}. \tag{61}$$

Hence, assembling the components of $T$ and then $|X(\eta,\xi,\phi)|^2$, and noting that all the terms in $\eta_0$ cancel in $T$ (the exponential terms because eq. (16b) shows that $\sigma I^{1/2}=2I/B$), yields

$$T = (2I/B) \cdot (8I/eF) \cdot \exp[-bI^{3/2}/F], \tag{62}$$

$$|X(\eta,\xi,\phi)|^2 = = \pi^{-1} \cdot (2I/B)^3 \cdot (eF\eta/2I-1)^{-1/2} \cdot \exp[-(2I/B)\xi] \cdot (8I/eF) \cdot \exp[-bI^{3/2}/F]. \tag{63}$$

To apply eq. (63) to the hydrogen atom, in the low field limit, using atomic units, put $e \to 1$, $B \to 1$, $I \to \frac{1}{2}$, $b \to 2^{5/2}/3$; this yields the expression given by LL as eq. (3) on p.257 of [18] (p. 275 of [26]), with the notational difference that it has been thought clearer to use $X(\eta,\xi,\phi)$ here rather than their symbol $\chi$.

### 3.4 Integration over all emission directions

In order to integrate over all emission directions (i.e. over $\phi$ and $\xi$), LL consider a plane normal to the

*z*-axis, well outside the barrier (so that 1/*r* can be treated as negligibly small). Their tactic is to formulate an expression for total emitted probability current *w* in cylindrical coordinates, and then convert this to parabolic coordinates before integrating.

Let $\rho$ denote radial distance in this plane, measured from its intersection with the *z*-axis. Using the relevant equation of continuity, the total probability current *w* crossing this plane can be found from

$$w = \int_0^{2\pi} \int_0^{\infty} |\Psi|^2 \, u_z(z,\rho,\phi) \cdot \rho \, d\rho \, d\phi , \tag{64}$$

where $u_z(z,\rho,\phi)$ is the component of electron velocity in the *z*-direction, at position $(z,\rho,\phi)$. The ESFI rate-constant $K_e$ is related to *w* by

$$w = \Pi K_e , \tag{65}$$

where $\Pi$ is the probability that the electron is "in the atom". Because $\Psi$ is normalised by setting $\Pi=1$, $K_e$ is numerically equal to *w*.

From eq. (11), when $\eta$ is large and $\xi$ small:

$$\rho^2 = x^2 + y^2 = \xi\eta; \quad \rho\,d\rho \approx \tfrac{1}{2}\eta\,d\xi . \tag{66}$$

Hence, on integrating with respect to $\phi$, and using eq. (49), eq. (64) yields

$$K_e = w = \pi \int_0^{\infty} |X|^2 \, u_z \, d\xi . \tag{67}$$

On the *z*-axis, at large distance from the nucleus, the kinetic energy of a classical point electron with classical velocity $u_a$ along the axis would be given by

$$\tfrac{1}{2}m_e u_a^2 \approx eFz - I \approx \tfrac{1}{2}eF\eta - I \; , \tag{68}$$

$$u_a \approx (2I/m_e)^{1/2}(eF\eta/2I - 1)^{1/2} \; . \tag{69}$$

Off-axis, because the electron path will not be radial but will be bent towards the $z$-axis, we have that $u_a\cos\theta < u_z(z,\rho,\phi) < u_a$, where $\theta$ is the angle shown in figure 1. When $\eta \gg \xi$, it can be shown that $\cos\theta \approx 1 - 2\xi/\eta \approx 1$. Since $\eta \gg \xi$, and since most of the contribution to integral (67) comes from relatively small values of $\rho$, $\theta$ and $\xi$, LL approximate $u_z \approx u_a$.

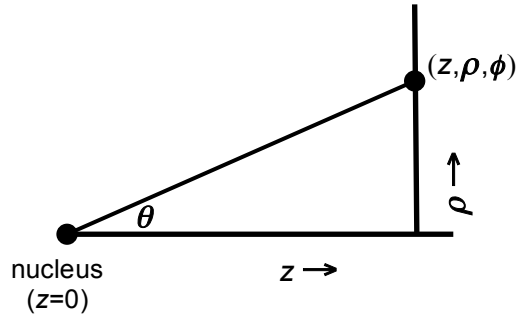

**Fig. 1.** To illustrate the definition of the angle $\theta$.

Inserting eq. (63) into eq. (67), and putting $u_z = u_a$, yields

$$K_e = (2I/B)^3 (2I/m_e)^{1/2} \cdot (8I/eF) \cdot \exp[-bI^{3/2}/F] \cdot \int_0^\infty \exp[-(2I/B)\xi]\,d\xi \tag{70}$$

$$K_e = (2I/B)^2 (2I/m_e)^{1/2} \cdot (8I/eF) \cdot \exp[-bI^{3/2}/F] \; . \tag{71}$$

On noting from eq. (16b) that $1/B^2 = \sigma^2/4I$, and using eq. (4), eq. (71) reduces to

$$K_e = C_{FI} \cdot (I^{5/2}/F) \cdot \exp[-bI^{3/2}/F] \; , \tag{72a}$$

where $C_{FI}$ is an universal constant, called here the *field ionization constant*, and given by

$$C_{FI} \equiv 2^{9/2} m_e^{1/2} / e\hbar^2 \approx 1.245\,354 \times 10^{17}\ \text{eV}^{-5/2}\ \text{V nm}^{-1}\ \text{s}^{-1}\ . \tag{72b}$$

Clearly, in atomic units $C_{FI} \rightarrow 2^{9/2}$.

Alternatively, since $I=Z^2 I_H$, eq. (72a) can be written in a form that explicitly involves $Z$, namely

$$K_e \approx C_{FI} Z^5 I_H^{5/2} F^{-1} \exp[-bZ^3 I_H^{3/2}/F]\ . \tag{73}$$

Appendix II shows that, for a hydrogen atom ($Z=1$), eq. (73) is equivalent to the Gaussian system equation given in Ref. [26].

To apply eq. (73) to a hydrogenic atom, using atomic units, put $C_{FI} \rightarrow 2^{9/2}$, $b \rightarrow 2^{5/2}/3$, $I_H \rightarrow I_{Hau}$, $F \rightarrow F_{au}$, where $I_{Hau}$ is the H-atom ionization energy, expressed in the atomic units system. This yields

$$K_e = Z^5 \cdot (2I_{Hau})^{5/2} \cdot (4/F_{au}) \cdot \exp[-Z^3 \cdot (2I_{Hau})^{3/2} \cdot 2/3F_{au}]\ . \tag{74}$$

Clearly, this reduces to the LL form for a hydrogen atom, eq. (2a) above, on setting $Z=1$ and $I_{Hau}=½$.

## 4. Alternative forms for the rate-constant formula

### 4.1 Form involving an effective escape probability

Obviously, by using eq. (25) ($v_Z = I/\pi\hbar$), eq. (72) may be written

$$K_e = v_Z D^{eff}\ , \tag{75}$$

where an "effective escape probability" $D^{\text{eff}}$ is defined by

$$D^{\text{eff}} \equiv K_e/\nu_Z = \pi\hbar C_{\text{FI}} \cdot (I^{3/2}/F) \cdot \exp[-bI^{3/2}/F] \tag{76}$$

(see Appendix I for the value of $\pi\hbar C_{\text{FI}}$). Note that $D^{\text{eff}}$ is not conceptually identical with the one-dimensional tunnelling probability $D^{\text{1d}}$ of eq. (1), because $D^{\text{eff}}$ takes three-dimensional effects into account.

Further, we can write

$$D^{\text{eff}} \equiv P_g \cdot T, \tag{77}$$

where $P_g$ is a dimensionless geometrical pre-factor, and use eqns (62) and (71), and then eqns (4), (16b) and (17), to obtain

$$P_g = K_e/T\nu_Z = (2I/B)(2I/m_e)^{1/2}/\nu_Z = 2\pi. \tag{78}$$

$$K_e = \nu_Z P_g T = 2\pi\nu_Z T = \omega_Z T, \tag{79}$$

where $\omega_Z$ is the classical angular frequency corresponding to $\nu_Z$.

## 4.2 Form involving a JWKB integral for barrier strength

In the physical discussion above, it is important that $\eta_0$ should be inside the barrier. However, terms in $\eta_0$ cancel out, both in LL's original derivation and above; thus—for evaluating the mathematics—the actual value of $\eta_0$ seems relatively unimportant. Hence, $\eta_0$ can be chosen to coincide with the inner zero of $\mathcal{M}(\eta)$, denoted here by $\eta_{in}$. The symbol $\eta_{out}$ [$\equiv \eta_1$] is now used to denote the outer zero.

Reconsider eqns (56c) and (56d), which can be replaced by

$$T = T_{JWKB} \equiv P_{JWKB} \exp[-\mathcal{G}^*] \equiv (2I/B)\eta_{in} \exp[-(2I/B)\eta_{in}] \cdot \exp[-\mathcal{G}^*], \tag{80a}$$

$$\mathcal{G}^* \equiv 2\sigma \int_{\eta_{in}}^{\eta_{out}} \mathcal{M}^{1/2}(\eta) d\eta, \tag{80b}$$

where $\mathcal{G}^*$ is a "transformed" barrier-strength expression defined by eq. (80a), and $P_{JWKB}$ [$=(2I/B)\eta_{in} \exp[-(2I/B)\eta_{in}]$ is a *JWKB-form pre-factor* defined via eq. (80a). Inserting expression (80a) for $T$ into eq. (79) yields

$$K_e = v_Z \cdot 2\pi P_{JWKB} \exp[-\mathcal{G}^*] \equiv v_Z P^{eff} \exp[-\mathcal{G}^*] = v_Z D^{eff} \tag{81}$$

where an *effective tunnelling pre-factor* $P^{eff}$ [$\equiv 2\pi P_{JWKB}$] is defined via eq. (81). This result, for ESFI, shows more clearly than some other treatments that (in non-planar geometries) $P^{eff}$ contains both a term relating to tunnelling along a characteristic path and a geometrical term relating to summation over all possible paths.

A more detailed approximate expression for $P^{eff}$ is obtained as follows. Equations (47) and (27) show that, in the limit of low field, $\eta_{in}$ is obtained by solving eq. (82) to give result (83):

$$I - B/\eta - 1/\sigma^2\eta^2 = I - B/\eta - B^2/4I\eta^2 = 0, \tag{82}$$

$$\eta_{in} \approx (B/2I)\left(1+\sqrt{2}\right). \tag{83}$$

Consequently, from eqns (80b) and (81), $P_{JWKB} \approx (1+\sqrt{2})\exp[-(1+\sqrt{2})]$, and $P^{eff}$ is given by

$$P^{\text{eff}} \approx 2\pi P_{\text{JWKB}} = 2\pi\left(1+\sqrt{2}\right)\exp\left[-\left(1+\sqrt{2}\right)\right] \approx 1.36. \tag{84}$$

As indicated above, eq. (84) is not a simple-JWKB approximation (which would have $P^{\text{eff}}=1$), but is an approximation that takes into account both: (a) that the simple-JWKB integral does not usually provide an exactly correct treatment of the quantum mechanics of tunnelling along a particular path (see [38]); and (b) effects associated with the three-dimensionality of the physical situation. This parameter $P^{\text{eff}}$ is conceptually more complex than the parameter $P_t$ in eq. (7).

**4.3 Equivalent result in a Cartesian context**

If eqns (47) and (80b), as evaluated along the symmetry axis, are converted to Cartesian coordinates, using $\eta=2z$, and we retain a modified typeface for the transformed motive energy (that results from the wave-function transformation (41)), then (for this path along the symmetry-axis)

$$\mathcal{G}^* = 2\sigma \int_{\eta_{\text{in}}}^{\eta_{\text{out}}} \mathfrak{M}^{1/2}(\eta)\,\mathrm{d}\eta = 2\sigma \int_{z_{\text{in}}}^{z_{\text{out}}} \mathfrak{M}^{1/2}(z)\,\mathrm{d}z, \tag{85}$$

where $z_{\text{in}}$ and $z_{\text{out}}$ are the zeros of the transformed motive energy

$$\mathfrak{M}(z) = I - eFz - Ze^2/8\pi\varepsilon_0 z - \hbar^2/8m_e z^2. \tag{86}$$

This expression $\mathfrak{M}(z)$ is not the one that would be derived by a naive one-dimensional consideration of energy terms taken along the $z$-axis, which would be

$$M^{\text{1D}}(z) = I - eFz - Ze^2/4\pi\varepsilon_0 z. \tag{87}$$

The physical meaning, and implications, of result (86) will be discussed elsewhere. Briefly, this result may imply that, if the Landau and Lifshitz approach to quasi-classical quantum mechanics is correct, and hence that transformation of the motive-energy expression is needed in at least some non-planar emitter geometries, then questions of principle arise about the validity of existing treatments of field electron emission in non-planar geometries.

5. Discussion

5.1 Summary

This paper has derived ISQ formulae for the rate-constant for ESFI of a hydrogenic atom in its ground electronic state, in the deep tunnelling regime. By defining a new universal constant $C_{FI}$, these formulae can be put in the ISQ short forms of eqns (72) and (73), which display the result (needed for future work) that the ionization energy appears in the rate-constant pre-exponential as $I^{5/2}$. At critical points in the derivation, it has been shown that ISQ formulae transform correctly to the related "atomics units" formulae that appear in LL's treatment.

The derivation uses constants ($I_H$, $a_0$) that relate to a hydrogen atom, constants that relate more generally to a hydrogenic atom ($Z$, $B$, $I$, $a_Z$, $v_Z$), and several defined universal constants ($\sigma$, $b$, $C_{FI}$). Expressions for these constants in terms of the fundamental constants are given in Appendix I. Where relevant, values for constants are given both in SI units and in the ISQ-compatible units often used in field electron and field ion emission; equivalent values in the atomic units system are also shown.

It has also been shown that the derived ISQ formula can be put into an "attempt frequency" form, as eq. (81). An expression has been given for the resulting effective tunnelling probability $D^{eff}$, and it has been shown that $D^{eff}$ can be written as $P^{eff}\exp[-\mathcal{G}^*]$, where $\mathcal{G}^*$ is a barrier-strength expression given by a (transformed) JWKB-type integral taken along the symmetry axis, and $P^{eff}$ is an effective tunnelling pre-factor (evaluated as about 1.36 for a hydrogen atom).

## 5.2 Comments

An advantage of the atomic units system is that formulae and proofs are usually more concise than ISQ formulae. The present ISQ proof is certainly longer and more involved than LL's atomic-units proof, but this is partly because, for transparency, the author has thought it helpful to discuss numerous small details. However, the resulting ISQ formulae, eqns. (72) and (73), are not significantly more complicated than the atomic-units formula, eq. (2a), and have the advantage that dependence on ionization energy or effective nuclear charge $Z$, as well as that on electrostatic field, can be explicitly shown.

For the hydrogenic-atom ionization energy, the power dependence in the pre-exponential found here, namely $I^{5/2}$, is the same as that stated or implied by Refs [29-31]. A power-dependence of this form $I^{5/2}$ was also demonstrated by a reviewer of an earlier version of this paper, by carrying out a scaling transformation on the Schrödinger equation.

Another past concern (e.g., [21]), particularly for applications to near-surface ESFI, has been the form of the power dependence on $F$ in the pre-exponential. For future work, it is of interest to note how an LL-type treatment produces this. The dependence $F^{-1}$ arises because the component $g_2$ of the quasi-classical integral can be written $g_2 = -\ln\{\eta_0^{-1}(8I/eF)\}$. This results from: (a) the binomial expansion generating eq. (57); (b) integration of the second term in eq. (57) to yield eq. (60); and (c) the binomial expansion of the term $(1-eF\eta_0/2I)^{1/2}$ in the second bracket in eq. (60). Both terms in the LL rate-constant formula arise from evaluations performed at $\eta_0$.

For the author's longer-term research aims, an issue unresolved here is the size of any approximation-errors in the LL treatment and hence in the present treatment, particularly for field values relevant to practical situations. Another unresolved issue is the physical interpretation of the difference between eqns (86) and (87). These topics will be addressed elsewhere.

It also remains to be explained precisely why more realistic treatments (e.g., Refs [32-34]) of atomic ESFI yield results different from the $I^{5/2}$-dependence found for the hydrogenic atom.

In conclusion, it is hoped that this ISQ derivation will be particularly useful for those working in areas (outside laser physics) that involve practical applications of free-space or near-surface ESFI, and also that it may be useful to students who first encounter ESFI theory in a teaching context or at the start of research careers.

I thank the University of Surrey for provision of facilities and Professor John Xanthakis, of the Technical University of Athens, for pointing out that eq. (2b) appears in the second English edition of the Landau and Lifshitz textbook, but not in the first.

# Appendix I: Definitions and values of constants

Table 1 below provides definitions of constants relevant to this treatment. Table 2 provides values of relevant fundamental and universal constants, both in SI units and in the ISQ-compatible units often used in field electron and ion emission. Table 2 also shows values of the equivalent constants in the atomic units system. The SI values of the fundamental constants were taken (in January 2015) from the on-line database maintained by the US National Institute of Standards and Technology (NIST), and are based on the 2010 CODATA values [41].

**Table 1.** Formulae for defined constants relevant to this paper. Symbols for the fundamental constants have the meanings that they usually have in the International System of Quantities. Note in particular that the elementary charge $e$ is measured in coulombs or dimensionally equivalent units.

| Name | Symbol | Derivation | ISQ expression |
|---|---|---|---|
| "Coulomb-law constant" | $B_H$ | - | $e^2/4\pi\varepsilon_0$ |
| "Coulomb-law constant for hydrogenic atom with nuclear charge $Ze$" | $B$ | $ZB_H$ | $Ze^2/4\pi\varepsilon_0$ |
| Bohr radius | $a_0$ | - | $4\pi\varepsilon_0\hbar^2/e^2 m_e$ |
| Bohr-type radius for hydrogenic atom | $a_Z$ | $a_0/Z$ | $4\pi\varepsilon_0\hbar^2/Ze^2 m_e$ |
| Ionization energy of H-atom ground state, in zero external field (using free-space electron mass) | $I_H$ | $e^2/8\pi\varepsilon_0 a_0$ | $e^4 m_e/32\pi^2\varepsilon_0^2\hbar^2$ |
| As immediately above, but for hydrogenic atom | $I$ | $Z^2 I_H$ | $Z^2 e^4 m_e/32\pi^2\varepsilon_0^2\hbar^2$ |
| Classical vibration frequency of electron in lowest Bohr orbit, for an H atom in zero external field | $\nu_0$ | $I_H/\pi\hbar$ | $e^4 m_e/32\pi^3\varepsilon_0^2\hbar^3$ |
| As immediately above, but for hydrogenic atom | $\nu_Z$ | $I/\pi\hbar$ or $Z^2\nu_0$ | $Z^2 e^4 m_e/32\pi^3\varepsilon_0^2\hbar^3$ |
| Classical angular frequency for electron in lowest Bohr orbit, for an H atom in zero external field. [Also, atomic-system unit for angular frequency.] | $\omega_0$ | $2\pi\nu_0$ | $e^4 m_e/16\pi^2\varepsilon_0^2\hbar^3$ |
| "Schrödinger equation constant for electron" (Fowler and Nordheim's constant "$\kappa$" [37]) | $\sigma$ | - | $(2m_e)^{1/2}/\hbar$ |
| Second Fowler-Nordheim constant | $b$ | $4\kappa/3e$ | $(4/3)(2m_e)^{1/2}/e\hbar$ |
| Field ionization constant | $C_{FI}$ | $2^{7/2}\sigma^2/em_e^{1/2}$ | $2^{9/2} m_e^{1/2}/e\hbar^2$ |
| "Constant in attempt-frequency-based formula" | - | $\pi\hbar C_{FI}$ | $2^{9/2}\pi m_e^{1/2}/e\hbar$ |

**Table 2.** Values of fundamental and universal constants relevant to this paper. In most cases, values are given both in SI units and in the ISQ-compatible (dimensionally equivalent) units often used in field electron and field ion emission. The values of equivalent constants in the atomic units system are also shown. The entry "n/u" ("not used") indicates that, in scientific practice, the constant concerned is unlikely to be found expressed in SI units. Numerical values are given to a precision of seven significant figures (eight for fundamental constants and closely related quantities). For consistency, the value given for the hydrogen ionization energy $I_H$ is the theoretical one deduced using the free-space electron mass. For meanings of symbols, see Table 1.

| Symbol | SI units | | atomic-level units based on eV | | Value of equivalent constant in atomic units system |
|---|---|---|---|---|---|
| | Numerical value | Units | Numerical value | Units | |
| eV | $1.602\,176\,6 \times 10^{-19}$ | J | 1 | eV | - |
| $e$ | $1.602\,176\,6 \times 10^{-19}$ | C | 1 | eV V$^{-1}$ | 1 |
| $m_e$ | $9.109\,382\,9 \times 10^{-31}$ | kg | $5.685\,630 \times 10^{-30}$ | eV nm$^{-2}$ s$^2$ | 1 |
| $\hbar$ | $1.054\,571\,7 \times 10^{-34}$ | J s | $6.582\,119 \times 10^{-16}$ | eV s | 1 |
| $\varepsilon_0$ | $8.854\,187\,8 \times 10^{-12}$ | F m$^{-1}$ | $5.526\,350 \times 10^{-2}$ | eV V$^{-2}$ nm$^{-1}$ | $1/4\pi$ |
| $4\pi\varepsilon_0$ | $1.112\,650\,1 \times 10^{-10}$ | F m$^{-1}$ | $0.694\,461\,6$ | eV V$^{-2}$ nm$^{-1}$ | 1 |
| $B_H$ | $2.899\,158\,9 \times 10^{-27}$ | J m | $1.439\,964$ | eV nm | 1 |
| $a_0$ | $5.291\,772 \times 10^{-11}$ | m | $5.291\,772 \times 10^{-2}$ | nm | 1 |
| $\nu_0$ | $6.579\,684 \times 10^{15}$ | Hz | $6.579\,684 \times 10^{15}$ | Hz | $1/2\pi$ |
| $\omega_0$ | $4.134\,137 \times 10^{16}$ | rad/s | $4.134\,137 \times 10^{16}$ | rad/s | 1 |
| $I_H$ | n/u | n/u | $13.605\,69$ | eV | 1/2 |
| $\sigma$ | n/u | n/u | $5.123\,167$ | eV$^{-1/2}$ nm$^{-1}$ | $2^{1/2}$ |
| $b$ | n/u | n/u | $6.830\,890$ | eV$^{-3/2}$ V nm$^{-1}$ | $2^{5/2}/3$ |
| $C_{FI}$ | n/u | n/u | $1.245\,354 \times 10^{17}$ | eV$^{-5/2}$ V nm$^{-1}$ s$^{-1}$ | $2^{9/2}$ |
| $\pi\hbar C_{FI}$ | n/u | n/u | $257.5185$ | eV$^{-3/2}$ V nm$^{-1}$ | $2^{9/2}\pi$ |

**Appendix II. Re-examination of the Landau and Lifshitz 1965 formula**

In the second English edition [26] of their textbook, Landau and Lifshitz gave a Gaussian-system expression for their rate-constant, reported above as eq. (2b). On substituting $e_s = e/(4\pi\varepsilon_0)^{1/2}$, $F_s = (4\pi\varepsilon_0)^{1/2} F$, the ISQ version of their formula is derived as

$$K_e = \left\{ 4m_e^3 e^9 / (4\pi\varepsilon_0)^5 \hbar^7 F \right\} \cdot \exp[-(2/3)m_e^2 e^5 / (4\pi\varepsilon_0)^3 \hbar^4 F]. \tag{A2.1}$$

It can be shown that this is an alternative (detailed) version of eq. (73), with $Z$=1. From Table 1, the detailed expression for $bI_H^{3/2}$ is the same as the coefficient of $1/F$ in the exponent of eq. (A2.1), and the detailed expression for $C_{FI}I_H^{5/2}$ is the same as the coefficient of $F^{-1}$ in the pre-exponential. Thus, in the case of the hydrogen atom, the result derived here is equivalent to the LL 1965 formula.

**Figure 1**

**Fig. 1.** To illustrate the definition of the angle $\theta$.

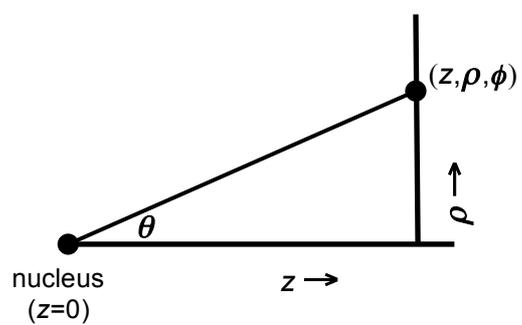